\documentclass[11pt,letterpaper]{article}

\usepackage{float}       
\usepackage{placeins}    
\usepackage{booktabs}    

\usepackage[margin=1in]{geometry}
\usepackage[utf8]{inputenc}
\usepackage[T1]{fontenc}
\usepackage{amsmath}
\usepackage{amsfonts}
\usepackage{amssymb}

\usepackage{graphicx}
\usepackage{subcaption}  

\usepackage[round]{natbib}

\usepackage{authblk}

\title{\bfseries \Large Weighting a Census as a Non-Probability Sample:\\
A Doubly Robust Framework for Correcting Differential\\
Undercoverage in Uruguay's 2023 Census}

\author[1,2]{Juan Pablo Ferreira\thanks{Corresponding author: \texttt{jferreir@ine.gub.uy} o \texttt{juanpablo.ferreira@fcea.edu.uy}}}
\author[1,3]{Juan Jos\'e Goyeneche\thanks{\texttt{jjgoye@fcea.edu.uy}}} 

\affil[1]{\small Methodology, Research, and Projects Division, Instituto Nacional de Estad\'istica, Uruguay}
\affil[2]{\small Facultad de Ciencias Econ\'omicas y de Administraci\'on, Universidad de la Rep\'ublica, Uruguay}
\affil[3]{\small Professor Emeritus, Facultad de Ciencias Econ\'omicas y de Administraci\'on, Universidad de la Rep\'ublica, Uruguay}

\date{\small May 2026}

\AtBeginDocument{
  \setlength{\abovedisplayskip}{10pt}       
  \setlength{\belowdisplayskip}{10pt}
  \setlength{\abovedisplayshortskip}{8pt}
  \setlength{\belowdisplayshortskip}{8pt}
}
\usepackage[colorlinks=true, allcolors=blue]{hyperref}
\begin{document}

\maketitle
\vspace{-2em} 

\begin{abstract}
The 2023 Uruguayan Census recorded a population of 3,444,451 with an estimated undercoverage of 10.3\%. Post-enumeration evidence shows that omission was non-random, concentrated in vulnerable areas, rural territories, and among young adults. Integrating administrative records (AR) recovered aggregate counts but did not resolve selection bias in outcome variables, as AR lack core census variables, exhibit urbanicity and institutional-visibility biases, and do not reconstruct households. Estimates derived from enumerated microdata remain biased. We treat effectively enumerated households as a non-probability sample with an unknown selection mechanism and construct weights using a doubly robust (DR) estimator. This framework combines a segment-level response-propensity model—using the web linkage rate as a contact proxy—with calibration to combined-census demographic totals (sex, age, department). Because the DR estimator is consistent when either model is correctly specified, it provides robustness against undercoverage misspecification. We describe the application at a scale of three million records, document its effect on social indicators, and present a variance approximation based on an equivalent stratified cluster design. Finally, we establish a methodological framework to guide national statistical offices on optimizing non-response adjustments based on their available registers and paradata.
\end{abstract}

\vspace{-0.5em} 
\noindent\textbf{Keywords:} doubly robust estimation; non-probability samples; census undercoverage; nonresponse weighting; calibration; response propensity.

\section{Introduction}

Census non-response has been rising across national statistical systems \citep{un2017}. The 2023 Uruguayan Population and Housing Census, conducted by the Instituto Nacional de Estad\'istica (INE), illustrates the consequences vividly. It enumerated a population of 3,444,451 against an estimated undercoverage of 10.3\% according to the Post-Enumeration Survey (PES), a marked break from Uruguay's historical benchmarks; the 2011 census, by comparison, had an estimated undercoverage of 4.07\%. An undercoverage of this magnitude poses inferential challenges that go well beyond the headline count.
\\

The central difficulty is not the level of undercoverage but its structure. Evidence from the PES shows that omission was concentrated in specific population segments rather than spread uniformly across the target population. In Montevideo, which holds roughly 40\% of the national population, the omission rate reached about 18\% in the lowest socioeconomic stratum against roughly 3\% in the highest. Rural omission was approximately double the national average, and age-sex patterns reveal the underrepresentation of young adults, with omission rates near 14\% to 15\% for those aged 25 to 35 and declining progressively in older cohorts. As a result, the enumerated population does not reproduce the demographic and socioeconomic structure of the target population: it is systematically older and better off than the population it is meant to describe.
\\

To recover the aggregate count, INE assembled a \emph{combined census} that integrated field enumeration with administrative records (AR) \citep{wallgren2014}. Using a selection algorithm based on administrative ``signs of life'' (i.e., individual-level interactions with state databases), approximately 350,000 non-enumerated persons were incorporated, consistent with the 10.3\% undercoverage rate. This procedure improves the population count and its age-sex profile. It does not, however, resolve the substantive problem. The administrative records carry only basic demographics (i.e., sex, age, and an approximate geographic reference), lack the roughly one hundred core census variables such as educational attainment and housing conditions, and were not used to reconstruct households, so most appended individuals remain unlinked to a residential unit. Knowledge of population totals does not, on its own, remove selection bias in outcome variables when response propensity is correlated with those variables \citep{sarndal2005}. Mass imputation of the missing variables is neither advisable nor feasible at this scale \citep{littlerubin2019,reiter2007}, and variable-specific weighting would fracture the internal consistency between household-level and person-level records.
\\

A separate weighting system applied to the \emph{effectively enumerated} households is therefore required, because those households are the only units with complete socioeconomic information and a defined household structure. The strategy adopted by INE treats the enumerated population as a non-probability sample, in which the probability of being enumerated is unknown and depends on observable and unobservable factors \citep{elliott2017,chen2020}, and corrects for differential undercoverage using a doubly robust (DR) estimator that combines a response-propensity model with a superpopulation model implied by calibration \citep{chambers2012,kang2007}.
\\

This article presents that framework as an applied case study. Our contribution is threefold. First, we set out a coherent conceptual treatment of a national census as a non-probability sample and motivate doubly robust estimation as the natural response to an unknown selection mechanism. Second, we describe a concrete, production-scale implementation, covering response-propensity modeling at the census-segment level, calibration to combined-census demographic totals, and the handling of edge cases such as certainty units and locality-level inconsistencies, for a microdata file exceeding three million records. Third, we document the effect of the framework on core census indicators and a variance approximation suitable for external users. We intend the case to be informative for other national statistical offices facing similar coverage pressures.
\\

The remainder of the article is organized as follows. Section~2 develops the conceptual framework: the nature of census undercoverage, the interpretation of the census as a non-probability sample, the role and limits of administrative records, and the shortcomings of traditional imputation. Section~3 describes the data and the empirical diagnosis of undercoverage. Section~4 presents the doubly robust weighting methodology. Section~5 reports the impact on census indicators, and Section~6 the variance estimation and quality measures. Section~7 states the assumptions and limitations, and Section~8 concludes. Operational guidance for microdata users (i.e., variance computation in R, the rounding procedure that preserves totals, and a glossary) is provided as supplementary material.

\section{Conceptual Framework}

\subsection{Undercoverage, Omission, and Non-Response}

Census undercoverage is the failure to enumerate individuals, households, or dwellings belonging to the target population during field operations, and it arises from territorial access difficulties, technological barriers, operational constraints, and household decisions about participation \citep{groves2002}. It is not a homogeneous phenomenon. We distinguish absent residents (i.e., dwellings identified but whose occupants could not be contacted within the enumeration period), omission proper (i.e., units neither identified nor enumerated, inferable only through external sources or evaluation instruments; \citealp{un2010}), and within-household omission (i.e., an enumerated household for which not all members are recorded). Within a finite-population framework these correspond to distinct sources of unit and item non-response.
\\

Census undercoverage becomes problematic precisely when it does not occur completely at random but concentrates in specific segments, generating coverage bias \citep{bethlehem2011}. Under a deterministic view, the bias of the enumerated mean $\bar{Y}_E$ can be written as the product of the undercoverage rate and the gap between the enumerated and omitted means,

\begin{equation}
\mathrm{B}(\bar{Y}_E) = W_M \,(\bar{Y}_E - \bar{Y}_M),
\end{equation}

so that bias depends not only on the magnitude of undercoverage but fundamentally on its differential nature. Equivalently, under a stochastic response model the bias can be expressed through the covariance between the enumeration probability $\phi$ and the variable of interest $y$,

\begin{equation}
\mathrm{B}(\bar{Y}_E) \approx \frac{\mathrm{cov}(\phi, y)}{\bar{\phi}},
\end{equation}

where $\bar{\phi}$ is the mean response propensity. Both expressions imply that, because 2023 undercoverage was concentrated in vulnerable strata, the term $(\bar{Y}_E - \bar{Y}_M)$ is large and the covariance is non-zero; the direct use of enumerated data without adjustment is therefore not defensible for differentially distributed variables. A further practical consequence is that absolute estimates (i.e., totals or counts) cannot be obtained for most indicators from the enumerated data alone, and the comparison of intercensal net change against the 2011 census (itself enumerated at a much lower undercoverage) is not meaningful without adjustment.

\begin{figure}[ht]
\centering
\includegraphics[width=0.8\textwidth]{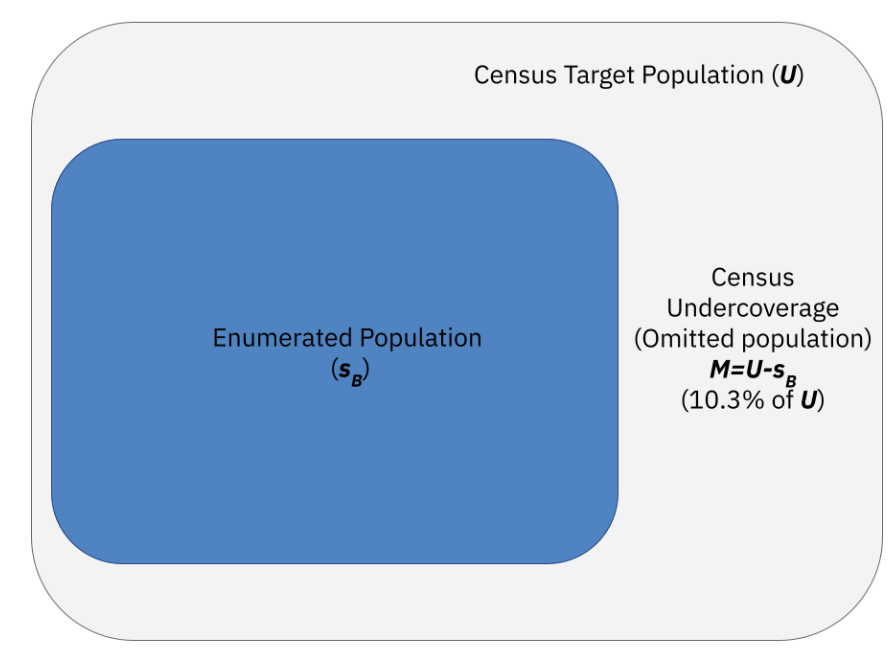}
\caption{Conceptual diagram of census undercoverage.}
\label{fig:conceptual}
\end{figure}

\subsection{The Census as a Non-Probability Sample}

When undercoverage is heterogeneous, the set of effectively enumerated households and individuals can be interpreted as a non-probability sample of the target population \citep{meng2018,elliott2017}. The selection mechanism does not follow a known sampling design but a complex response-generation process driven by observable and unobservable factors: the socio-demographic characteristics of the household and external conditions such as fieldwork operational constraints. Because the selection mechanism is unknown, design-based inference cannot by itself guarantee unbiased estimates, which motivates the use of methods developed for non-probability data sources, where the selection process is modeled and auxiliary information is used to mitigate bias \citep{chen2020}.

\subsection{Administrative Records and the Limits of Count-Based Integration}

Administrative records are a valuable auxiliary source for improving population counts, and their integration into the combined census materially improved the aggregate figures by sex, age, and department. Their contribution to the correction of substantive bias is, however, limited, for reasons that are structural rather than incidental.
\\

First, AR coverage is conditional on an individual's interaction with state institutions, which induces an \emph{institutional-visibility bias}: socioeconomically vulnerable and informally employed individuals are more likely to be absent from formal registries and thus to be misclassified \citep{wallgren2014,zhang2012}. Second, the residential-selection and localization models used to integrate AR were trained against a census whose coverage was itself non-random; using ``censused status'' as a target variable risks reproducing rather than correcting the field bias \citep{hand2006}. Third, reliance on administrative anchors generates an \emph{urbanicity bias}: in areas with irregular settlement or dispersed rurality, where such anchors are weak, individuals tend to be misallocated toward denser urban centers, deflating small localities and rural areas. Fourth, because the integration operates at the individual level, it does not reconstruct the household (i.e., the primary unit of census analysis), breaking the person-to-dwelling linkage for a substantial share of the appended population \citep{statsnz2019} and precluding the computation of household indicators such as overcrowding.
\\

These observations are not a criticism of the combined census, which remains an essential input to the weighting stage; they delimit what count-based integration can and cannot accomplish. It corrects imbalances in aggregate counts. It does not, and cannot, substitute for the missing substantive variables or correct the selection bias in outcome variables. For that purpose the administrative integration algorithm is best regarded as a record-classification tool, and a unified weighting system as the methodologically appropriate route to restoring representativeness. Consistent with this delimitation, INE used combined-census aggregates as calibration benchmarks only at levels of aggregation where their quality was judged sufficient (i.e., age-sex by department), deliberately avoiding the finer geographic scales most affected by allocation error.

\subsection{Limitations of Traditional Imputation}

Undercoverage, and in particular absent residents, has traditionally been treated through hot-deck imputation, which fills each missing record with the values of a donor unit drawn from a pool of enumerated households sharing observed auxiliary characteristics \citep{fellegi1976}. This preserves internal consistency but cannot correct coverage bias when undercoverage is non-random, because donors reproduce the distribution of the observed population rather than that of the missing one. The method implicitly assumes that, conditional on the donor-pool variables, observed units are representative of unobserved ones, i.e., a missing-at-random assumption that fails under differential undercoverage.
\\

INE evaluated this pathway empirically during the 2023 round by comparing estimates from the enumerated households alone with estimates that incorporated imputed absent-resident records. The two sets of estimates were substantially similar, confirming that imputation did not mitigate undercoverage bias in this context. This diagnosis motivated the move to a non-probability-sample treatment of the enumerated population, integrating methods designed for that setting to reduce estimation bias while preserving the analytic value of the census microdata.

\subsection{Methodological Decision Framework: Generalizability and Applicability}

To guide National Statistical Offices (NSOs) and survey methodologists in determining the operational applicability of this weighting framework, a systematic decision matrix is established based on the latent missingness mechanism and the structural quality of the available statistical infrastructure.

\noindent\textbf{Scenario A: Data-scarce environments and ignorable non-response.} \\
When census omission is assumed to follow a Missing Completely at Random (MCAR) mechanism, the probability of response is orthogonal to both observed and unobserved characteristics. Empirical demographic literature, however, demonstrates that the MCAR assumption is virtually impossible to satisfy in modern census operations, as non-response is systematically correlated with socio-economic status and geographic accessibility. Under severe data scarcity (e.g., when the NSO lacks integrated administrative registries or high-coverage validation surveys), traditional imputation techniques such as Hot-Deck are often utilized as a constrained operational recourse. To minimize selection bias under these non-ignorable conditions, the conventional Hot-Deck implementation must be modified toward a localized neighborhood framework. Specifically, the donor pool should be restricted to adjacent spatial units exhibiting high resistance to enumeration (e.g., effectively enumerated households within the same micro-territory requiring the maximum number of field protocol visits). This ensures that the imputed units reflect the latent socio-economic profiles of the omitted households in the absence of administrative auxiliary data.

\noindent\textbf{Scenario B: Differential non-response with fragmented infrastructure.} \\
If census omission follows a Missing Not at Random (MNAR) pattern, where undercoverage is highly differential and concentrated within hard-to-reach sub-populations, the methodological strategy depends entirely on the auxiliary data infrastructure. If administrative records are heavily fragmented or unavailable across public institutions, the NSO must rely on a high-coverage Post-Enumeration Survey (PES) as the primary auxiliary source to implement Inverse Probability Weighting (IPW) adjustments. Under this scenario, the NSO must commit to an expanded, large-scale PES sampling design capable of producing reliable omission rate estimates for small analytical domains. Without a sufficiently powered PES to model these localized propensities, estimation is restricted to basic post-stratification or simple IPW based on limited census paradata, introducing a substantial risk of collapsing domains and generating severe residual selection bias.

\noindent\textbf{Scenario C: Observable non-response with integrated statistical registers.} \\
The application of the Doubly Robust (DR) framework developed in this study becomes methodologically mandatory when census omission operates under a Missing at Random (MAR) or MNAR mechanism conditional on observables, and the statistical ecosystem provides access to high-coverage integrated administrative records containing comprehensive demographic and geographic predictors. By combining a response propensity sub-model with an outcome prediction sub-model, the DR estimator establishes an essential mathematical safeguard against model misspecification. Under these data conditions, the estimator guarantees consistency if at least one of the two auxiliary sub-models is correctly specified, offering a critical protection against estimation errors that traditional single-stage weighting adjustments cannot provide.

\noindent\textbf{Cross-Cutting Calibration: Demographic Projections as Universal Benchmarks.} \\
Crucially, the operationalization of the second-stage calibration within the DR framework is independent of the microdata availability discussed across the preceding scenarios. In data-scarce or fragmented environments where individual-level administrative registers are missing (i.e., Scenarios A and B), the vector of population totals $\mathbf{X}$ can be successfully sourced from aggregate demographic estimates. By projecting a baseline population through the demographic balancing equation under a cohort-component method—accounting for births, deaths, and net migration—NSOs can derive structural control totals for key marginal domains (e.g., sex, age groups, and geographical regions). 

Furthermore, if official demographic projections are restricted to broader, aggregated age brackets, indirect estimation and demographic interpolation techniques—such as Sprague multipliers or monotonic cubic splines—can be dynamically deployed to reconstruct single-year-of-age benchmarks. Benchmarking the survey weights to these demographic projections mathematically forces the final aggregated weights to recover the country's official population trajectory ($\sum w_i = \hat{N}$), ensuring macro-demographic consistency regardless of the primary data collection infrastructure.

Additionally, when an NSO deploys a mixed-mode data collection strategy (e.g., combining self-administered web questionnaires with subsequent face-to-face interviews), the generated paradata offers a critical methodological advantage. Operational metrics, such as contact and completion rates across modes, can be modeled to compute proxy estimates of omission propensities during the field operation. These proxies allow the NSO to dynamically capture the behavioral mechanisms behind non-response, serving as a powerful auxiliary input to adjust weighting structures even before formal coverage surveys or administrative register linkages are completed.

\section{Data and Empirical Diagnosis}

\subsection{The Census Operation and the Post-Enumeration Survey}

The 2023 census used a mixed-mode design combining computer-assisted personal interviewing (CAPI) and a self-administered web instrument (CAWI). Regardless of whether a household first responded by web, it had to be contacted in person by an enumerator to verify the address and link the questionnaire to the corresponding dwelling, a feature we exploit below.
\\

The primary instrument for diagnosing undercoverage is the 2023 National Census Evaluation Survey (PES), designed to estimate fieldwork coverage and characterize omission patterns at the national level and for broad domains by territory, socioeconomic status, age, and sex. The PES confirms a clear socioeconomic gradient (i.e., omission rising as area-level socioeconomic status falls) and an age structure in which young adults are markedly underrepresented. These results provide the empirical basis for the adjustments described in Section~4 and serve as an external check on the model-based estimates.

\begin{figure}[ht]
\centering
\includegraphics[width=0.7\textwidth]{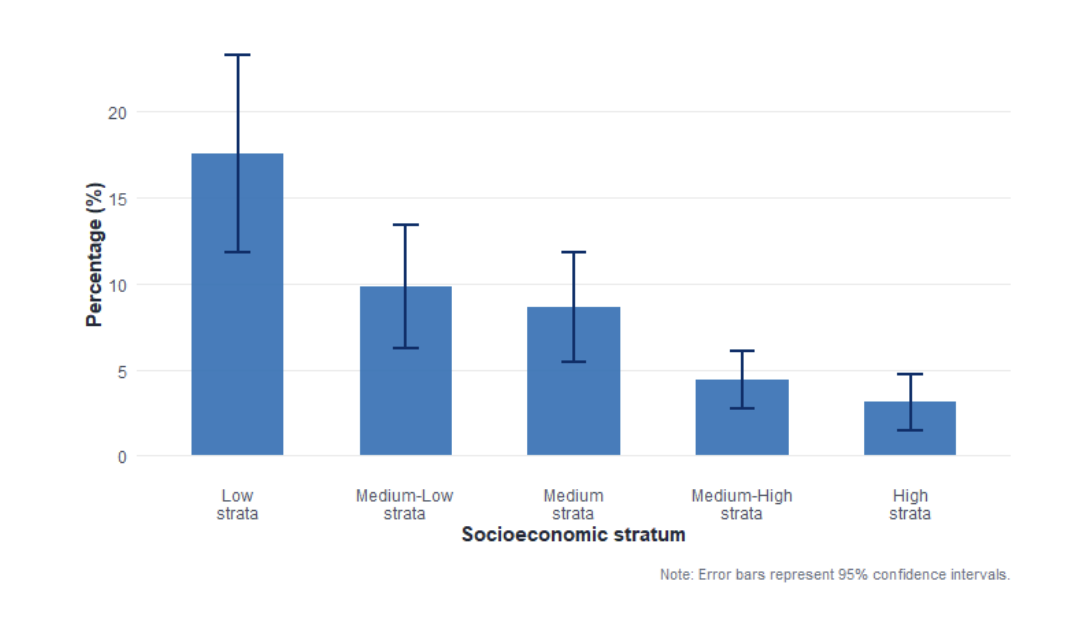}
\caption{Undercoverage estimates by socioeconomic stratum, Montevideo (95\% confidence intervals).}
\label{fig:ses}
\end{figure}

\begin{figure}[ht]
\centering
\includegraphics[width=0.7\textwidth]{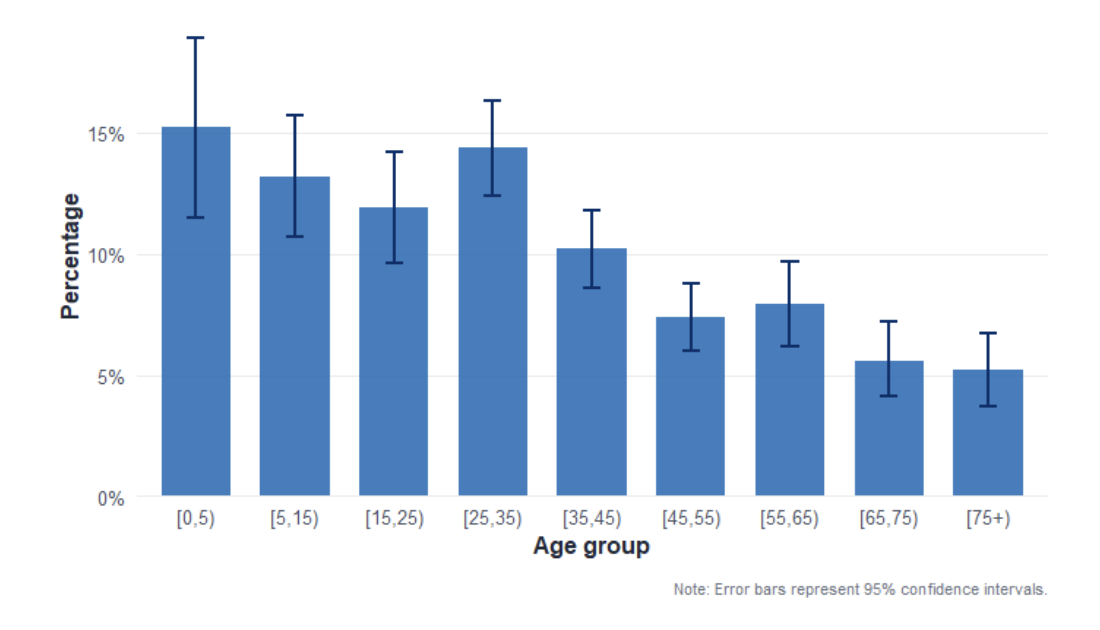}
\caption{Undercoverage estimates by age group (95\% confidence intervals).}
\label{fig:age}
\end{figure}

\FloatBarrier 

\subsection{Segment-Level Undercoverage: A Linkage-Rate Response Model}

The PES supports reliable estimates for broad domains but is not sized to deliver precise estimates at the census-segment level, the granularity at which the non-response adjustment operates. In the Uruguayan census cartography, the segment is the baseline enumeration area: it corresponds to groups of blocks in urban zones and to specific territorial extensions in dispersed rural areas. The 2023 cartography comprises approximately 4,500 such segments, and modeling omission across these micro-territorial units provides an ultra-granular spatial scale that captures localized socioeconomic heterogeneity invisible to broad domains. To estimate response rates at this scale we adopt a Linkage Rate-Based Census Response Model (LRBM).
\\

The LRBM rests on the decomposition of survey participation into contact and cooperation \citep{groves1998}, under which the response probability factors as
\begin{equation}
P(R) = P(C)\,P(K \mid C),
\end{equation}
with $P(C)$ the probability of establishing contact and $P(K \mid C)$ the probability of cooperation once contacted. Operational evidence and the mandatory nature of the census imply a low refusal rate, so cooperation is treated as near-universal, $P(K \mid C) \approx 1$, and the response probability is governed largely by the contact mechanism, $P(R) \approx P(C)$. The central assumption of the LRBM is that, within a census segment, the web-questionnaire linkage rate among CAWI households is an adequate proxy for the contact probability, and hence for the response rate, of households reached only in person: segments with lower linkage rates are taken to face greater contact difficulty and higher omission. This proxy was validated by comparing the model estimates with PES estimates at the national level and for broad domains, with high concordance between the two. The LRBM is used only to construct non-response adjustment factors, never to produce estimates directly.
\\

To ensure numerical stability, the framework enforces a minimum case threshold per unit. Where the sample size within a given segment is insufficient to yield stable estimates, the omission rate is computed by pooling contiguous neighboring segments that share administrative boundaries and homogeneous structural characteristics, thereby stabilizing the segment-level parameters. This geographic resolution enables precise adjustments that aggregated estimates, such as those from the PES, cannot deliver \citep{pfeffermann2013}, supporting the representativeness and precision of the final census parameters.
\\

To model the registration propensities at a finer geographical resolution, we exploit the census linkage rate at the segment level. This approach allows us to capture the sub-municipal heterogeneity that global administrative registers often mask.
\\

\begin{figure}[ht]
     \centering
     \begin{subfigure}[b]{0.48\textwidth}
         \centering
         \includegraphics[width=\textwidth]{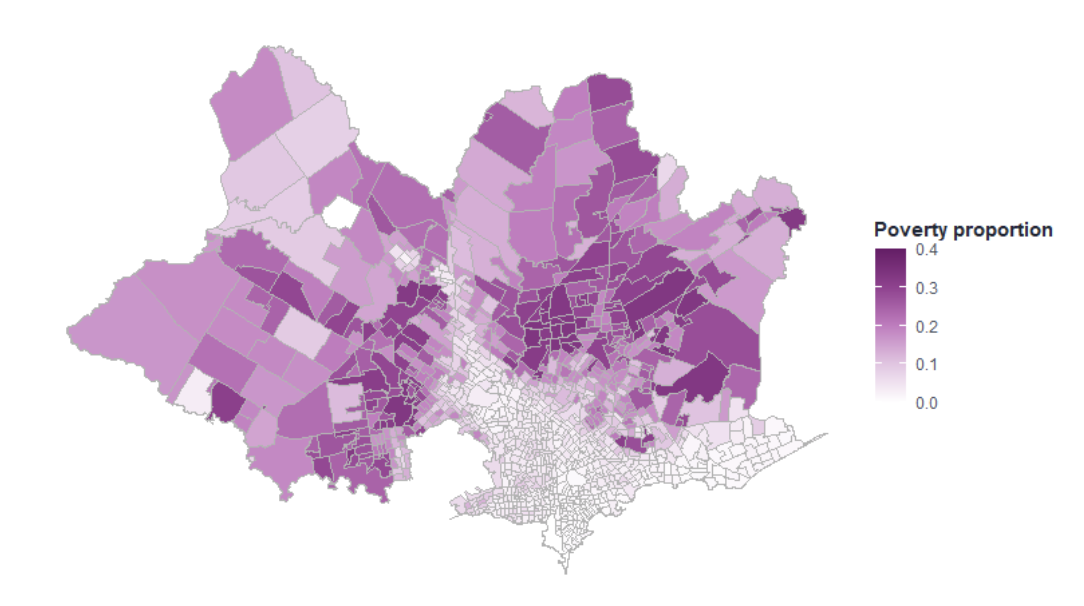}
         \caption{Census undercoverage (LRBM).}
         \label{fig:lrbm}
     \end{subfigure}
     \hfill
     \begin{subfigure}[b]{0.48\textwidth}
         \centering
         \includegraphics[width=\textwidth]{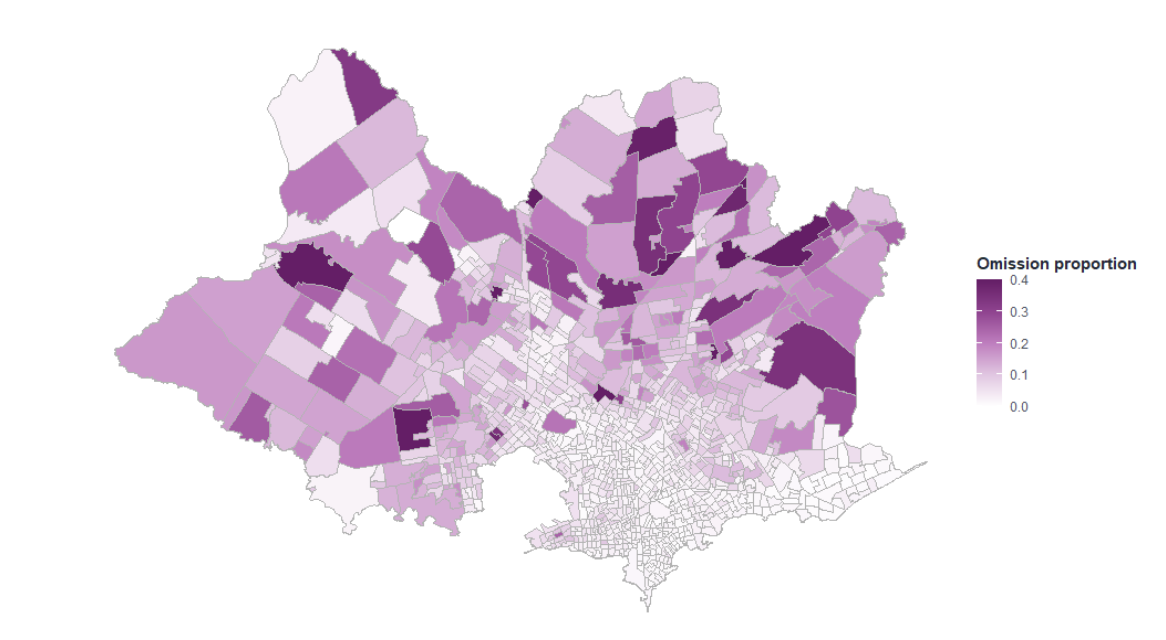}
         \caption{Poverty estimates (Fay-Herriot).}
         \label{fig:poverty}
     \end{subfigure}
     \caption{Segment-level estimates in Montevideo: comparison between estimated census undercoverage and poverty levels.}
     \label{fig:maps_comparison}
\end{figure}

Concerning the spatial distribution of non-response, a clear pattern emerges when analyzing the census segments. As illustrated in Figure~\ref{fig:lrbm}, undercoverage is not geographically uniform. The eastern and coastal areas of Montevideo display high linkage rates, whereas the northern and northwestern peripheries—traditionally characterized by lower income levels—exhibit a severe concentration of omitted households. To confirm this structural bias, Figure~\ref{fig:maps_comparison}b contrasts these results with the poverty estimates derived from the Fay-Herriot model.
\\

Undercoverage estimated through the LRBM overlaps spatially with small-area poverty estimated through a Fay-Herriot model: the correlation between the two is 0.67 ($p < 0.001$), and a regression of undercoverage on poverty has a positive slope of 0.47. The spatial congruence supports the interpretation of undercoverage as systematically linked to vulnerability \citep{groves2006}.
\\

Across the country, the distribution of segment-level omission is strongly right-skewed and varies markedly between departments.

\begin{figure}[H] 
\centering
\includegraphics[width=0.65\textwidth]{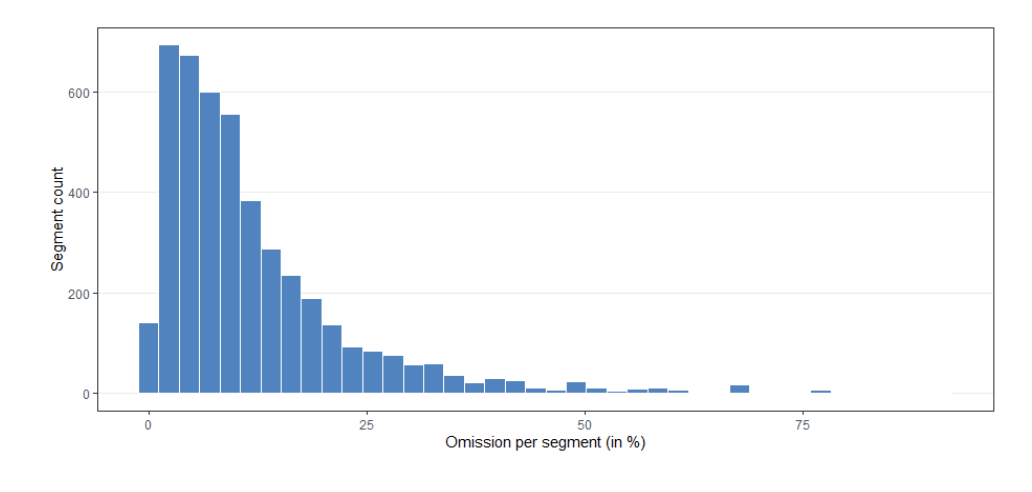} 
\vspace{-4pt} 
\caption{Distribution of segment-level undercoverage estimates.}
\label{fig:dist-dept}
\end{figure}

\subsection{Calibration Benchmarks and Auxiliary Variables}

Calibration uses population totals from the combined census by sex, single year of age, and department of residence; within the weighting framework these are treated as known totals free of sampling error. For response-propensity modeling, the auxiliary variables are the census segment, the data-collection mode (i.e., CAWI or CAPI), and the within-segment web linkage rate. Finer residential covariates (i.e., locality or neighborhood) were available but excluded from calibration because the administrative allocation algorithm misallocated appended individuals toward larger urban centers, producing implausible locality counts (e.g., combined-census undercoverage below 4\% in scattered rural areas where the PES and operational evidence indicate rates of roughly 20\%). Calibration was therefore restricted to aggregation levels where the benchmarks were sufficiently consistent, and a separate locality-level correction (Section~4.6) was introduced.

\nopagebreak 
\begin{figure}[H] 
\centering
\includegraphics[width=0.62\textwidth]{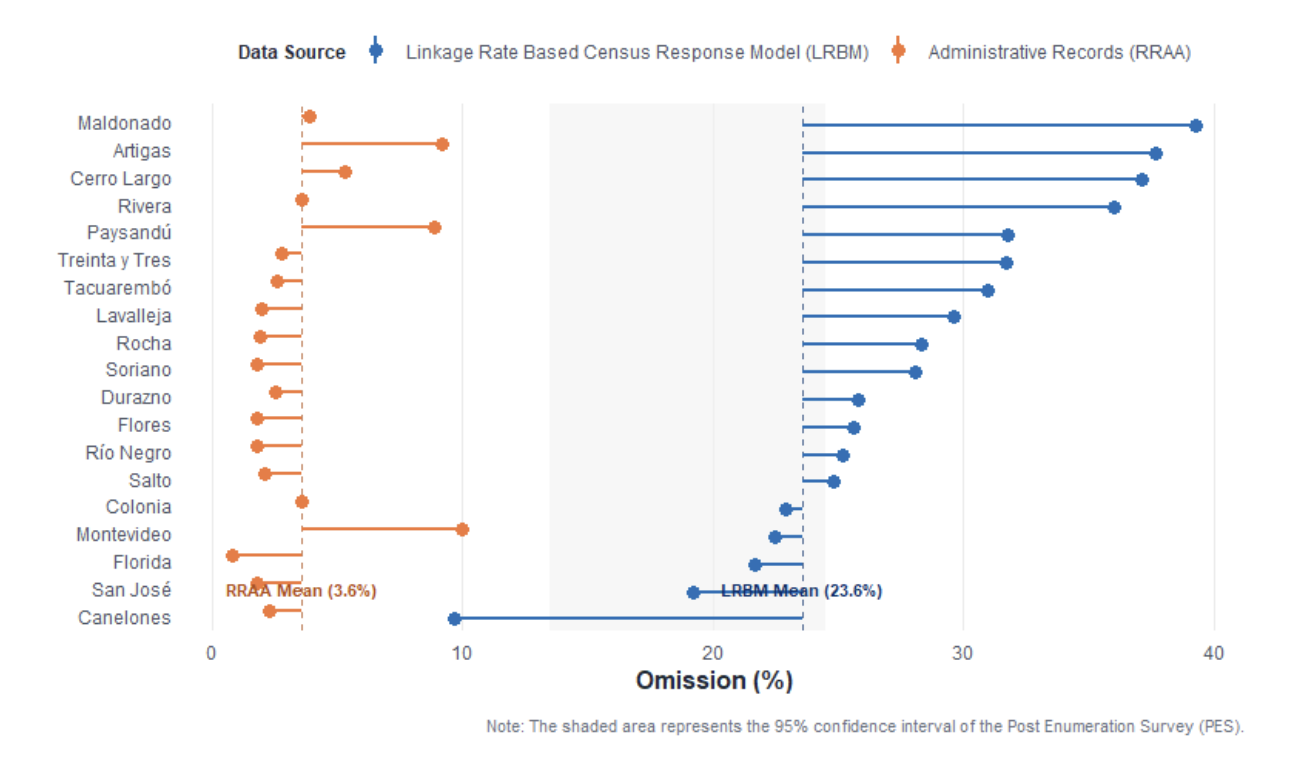} 
\vspace{-6pt} 
\caption{Census undercoverage in rural areas by department, by data source (LRBM and administrative records), compared with the PES confidence interval.}[cite: 3]
\label{fig:dept-source}[cite: 3]
\end{figure}

\section{A Doubly Robust Weighting Framework}

\subsection{Setup and Notation}

Let $U$ be the target population of size $N$ and $s_B$ the set of effectively enumerated units, interpreted as a non-probability sample. For unit $i$ let $y_i$ be the variable of interest, with population total $Y = \sum_{i \in U} y_i$ the parameter of interest. Let $R_i = 1$ if unit $i$ was enumerated and $R_i = 0$ otherwise. We assume the missing-at-random (MAR) condition \citep{rubin1976,littlerubin2019}, namely that response depends only on observed auxiliary variables,
\begin{equation}
P(R_i = 1 \mid y_i, \mathbf{x}_i) = P(R_i = 1 \mid \mathbf{x}_i),
\end{equation}
and define the response propensity $\phi_i = P(R_i = 1 \mid \mathbf{x}_i)$, where $\mathbf{x}_i$ is a vector of auxiliary covariates available for all units or defined groups (i.e., age, sex, geography, and operational variables).

\subsection{Inverse-Probability Weighting and the Model-Based Estimator}

If the propensities were known, the total could be estimated by inverse-probability weighting (IPW), an extension of the Horvitz-Thompson estimator that adjusts inclusion probabilities for non-response \citep{sarndal1992,seaman2013}. In practice $\phi_i$ is replaced by an estimate $\hat{\phi}_i$ from an assumed response model, giving
\begin{equation}
\hat{Y}_{\mathrm{IPW}} = \sum_{i \in s_B} \frac{y_i}{\hat{\phi}_i} = \sum_{i \in s_B} w_i^{nr}\, y_i .
\end{equation}
This estimator reduces bias when the response model is correctly specified but can be unstable if it is not. Alternatively, a superpopulation (working) model relates the outcome to the covariates, $E(y_i \mid \mathbf{x}_i) = m(\mathbf{x}_i)$, yielding the model-based estimator

\begin{equation}
\hat{Y}_{\mathrm{MB}} = \sum_{i \in s_B} y_i + \sum_{i \in U \setminus s_B} \hat{m}(\mathbf{x}_i),
\end{equation}

whose validity depends on the model estimated from $s_B$ remaining valid for the unobserved units \citep{valliant2000,chambers2012}.

\subsection{The Doubly Robust Estimator}

The doubly robust estimator combines the two \citep{chen2020}:
\begin{equation}
\hat{Y}_{\mathrm{DR}} = \sum_{i \in s_B} \frac{y_i - \hat{m}(\mathbf{x}_i)}{\hat{\phi}_i} + \sum_{i \in U} \hat{m}(\mathbf{x}_i).
\end{equation}

It is consistent if \emph{either} the response model or the superpopulation model is correctly specified \citep{kang2007}, a property of obvious value when the undercoverage mechanism is unknown. Its bias can be approximated as

\begin{equation}
\mathrm{B}(\hat{Y}_{\mathrm{DR}}) \approx \sum_{i \in U} \left(1 - \frac{R_i}{\phi_i}\right)\bigl(y_i - m(\mathbf{x}_i)\bigr),
\end{equation}

so that, in expectation, the systematic error is a product of the specification errors of the two models,
\begin{equation}
E\bigl[\mathrm{B}(\hat{Y}_{\mathrm{DR}})\bigr] \approx \sum_{i \in U} E\!\left[\frac{\phi_i}{\hat{\phi}_i} - 1\right] E\bigl[y_i - m(\mathbf{x}_i)\bigr],
\end{equation}

and vanishes if either $\hat{\phi}_i = \phi_i$ or $m(\mathbf{x}_i) = E[y_i]$. When the working model is linear, $m(\mathbf{x}_i) = \mathbf{x}_i^{\top}\boldsymbol{\beta}$, the DR estimator can be written as a weighted sum $\hat{Y}_{\mathrm{DR}} = \sum_{i \in s_B} w_i y_i$ with $w_i = w_i^{nr} g_i$ and

\begin{equation}
g_i = 1 + (\mathbf{X} - \hat{\mathbf{X}}_{\mathrm{IPW}})^{\top}\Bigl(\sum_{i \in s_B} w_i^{nr}\,\mathbf{x}_i \mathbf{x}_i^{\top}\Bigr)^{-1}\mathbf{x}_i,
\end{equation}

where $\mathbf{X} = \sum_{i \in U} \mathbf{x}_i$ and $\hat{\mathbf{X}}_{\mathrm{IPW}} = \sum_{i \in s_B} w_i^{nr}\mathbf{x}_i$.

The resulting weights satisfy the calibration (benchmarking) equations $\sum_{i \in s_B} w_i \mathbf{x}_i = \mathbf{X}$. Thus, under a linear working model the DR estimator coincides with a calibration estimator \citep{deville1992}: the two frameworks differ in derivation, since calibration imposes the constraints directly while the DR weights arise from combining an outcome model with a response model, yet they yield identical estimators. We exploit this equivalence by implementing the second stage as calibration.

\subsection{Stage One: Response-Propensity Adjustment}

The first stage models the probability that a household was enumerated. Because undercoverage manifests mainly as total household non-response or within-household omission, modeling is performed at the household level and all members of a household share its propensity and adjustment factor.
\\

Non-response adjustment classes are defined by census segments, on the assumption that grouping units into classes with similar response probabilities reduces non-response bias and that segment membership also correlates with the substantive variables \citep{little1986,littlevart2005}. Segments cluster households with relatively homogeneous geographic, socioeconomic, and accessibility characteristics, which supports both assumptions \citep{bethlehem2011}. Writing $c = 1, \dots, C$ for the classes, the propensity of household $i$ in segment $c$ is modeled as $\phi_i = \phi_c$ for all $i \in c$ \citep{valliant2018}. Using segments as classes targets the bias-variance trade-off explicitly: the variance of the weights,

\begin{equation}
V(w^{nr}) = \frac{1}{N}\sum_{c=1}^{C} N_c\left(\frac{1}{\hat{\phi}_c} - 1\right),
\end{equation}

with $N_c$ the segment size and $\hat{\phi}_c$ the estimated propensity, is minimized by maximizing intra-class homogeneity, and motivates an upper bound on the adjustment factors to prevent very small propensities from inflating variance.
\\

The non-response adjusted weight is $w_i^{nr} = 1/\hat{\phi}_i$, with $\hat{\phi}_i = 1 - \hat{W}_{M,c}$ and $\hat{W}_{M,c}$ the estimated undercoverage rate in segment $c$. These weights inflate units in higher-omission segments and are capped at a maximum of 2.5 to control variance \citep{valliant2018}. By construction their sum recovers the estimated census population, $\hat{N} = \sum_{i \in s_B} w_i^{nr} \approx N$.
\\

This stage corrects spatial bias associated with localized living conditions but does not, on its own, reproduce the demographic structure: a comparison of the age distribution of the enumerated population, the non-response adjusted estimates, and the combined census shows residual age-related imbalance. A calibration stage is therefore required.

\begin{figure}[ht]
\centering
\includegraphics[width=0.7\textwidth]{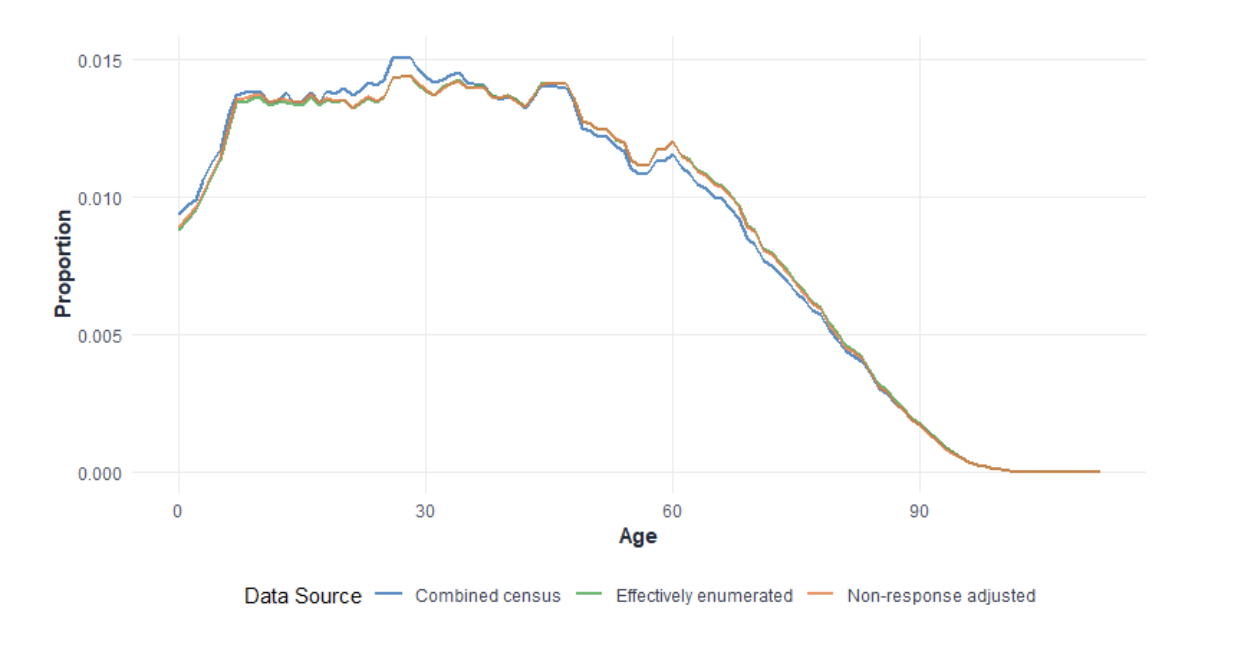}
\caption{Age distribution across data sources: enumerated population, non-response adjusted estimates, and combined census.}
\label{fig:age-sources}
\end{figure}

\subsection{Stage Two: Calibration to Combined-Census Totals}

Calibration adjusts the non-response weights so that aggregate estimates reproduce known population totals by sex, single year of age, and department. The calibration vector $\mathbf{x}_i$ consists of indicators of membership in these demographic groups, and under a linear distance the procedure is a generalized regression (post-stratification) estimator with implied working model

\begin{equation}
E_m(y_i) = \mathbf{x}_i^{\top}\boldsymbol{\beta} = \mu + \mathrm{age}_j + \mathrm{sex}_k + \mathrm{dept}_l + (\mathrm{age} \times \mathrm{sex} \times \mathrm{dept})_{jkl}.
\end{equation}

Letting $w_i^{nr}$ be the input weight, calibration finds $w_i = g_i w_i^{nr}$ as close as possible to $w_i^{nr}$ subject to $\sum_{i \in s_B} w_i \mathbf{x}_i = \mathbf{X}$, by minimizing a distance between calibrated and input weights \citep{deville1992}. When the calibration variables are associated with both the response mechanism and the outcomes, the resulting estimator inherits the doubly robust property; the assumption that age, sex, and department carry information about the substantive variables is reasonable given their well-known association with most social and economic characteristics.
\\

Because all members of a household must share a weight, the auxiliary information, though defined at the individual level, is incorporated through the integrated weighting method, which aggregates and averages the individual-level controls at the household level before calibration \citep{lemaitre1987,alexander1987}. This preserves household-person consistency at the cost of a slight increase in weight variability, and hence in the Kish design effect \citep{kish1965}, relative to allowing distinct individual weights. 
\\

\begin{figure}[H]
\centering
\includegraphics[width=0.60\textwidth]{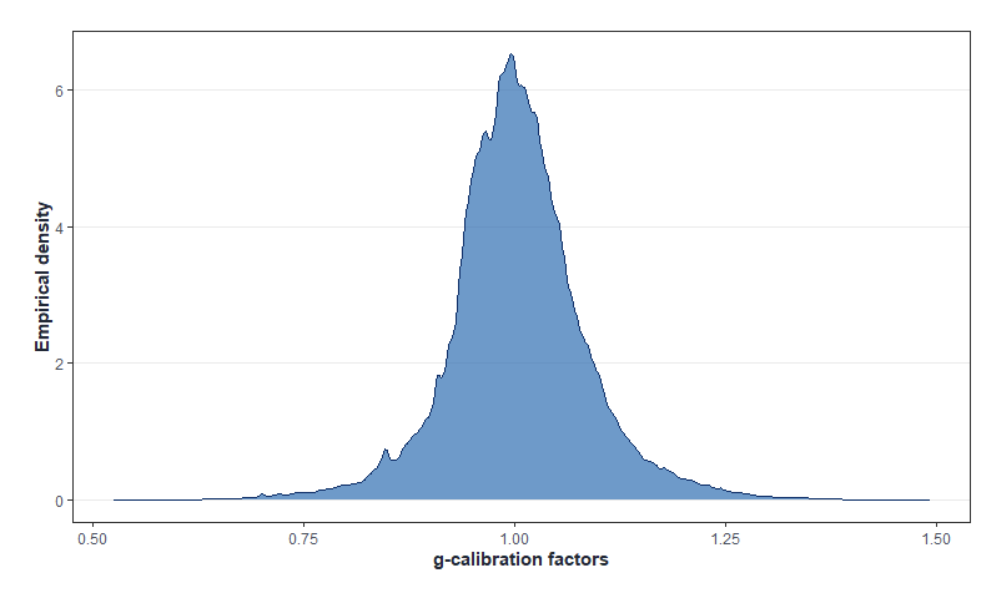}
\caption{Distribution of the calibration $g$-factors.}
\label{fig:gfactors}
\end{figure}

The empirical distribution of the resulting calibration multipliers provides direct evidence regarding the magnitude of these consistent adjustments. As visualised in Figure~\ref{fig:gfactors}, the global calibration $g$-factors are distributed symmetrically around one, demonstrating that the linear optimization constraints operate smoothly without forcing extreme scale distortions for the baseline sample.

\begin{figure}[H]
\centering
\includegraphics[width=0.65\textwidth]{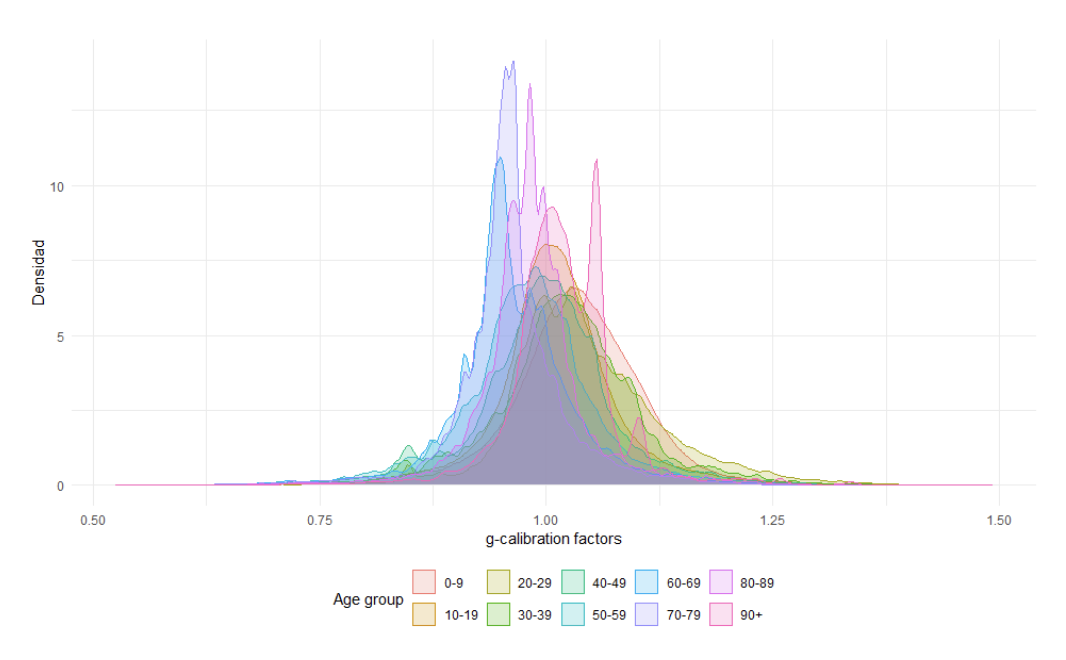}
\caption{Distribution of the calibration $g$-factors by age group.}
\label{fig:gfactors-age}
\end{figure}

However, a macro-level evaluation hides crucial structural variations across demographic strata. When these multipliers are disaggregated by specific cohorts, as plotted in Figure~\ref{fig:gfactors-age}, a somewhat heavier dispersion emerges within the intermediate age groups (i.e., young adults aged 25 to 35). This higher variance indicates that the calibration equations exerted a stronger correction pressure precisely on the specific demographic cells that had been most severely affected by differential census omission during field operations.

\begin{figure}[H]
\centering
\includegraphics[width=0.60\textwidth]{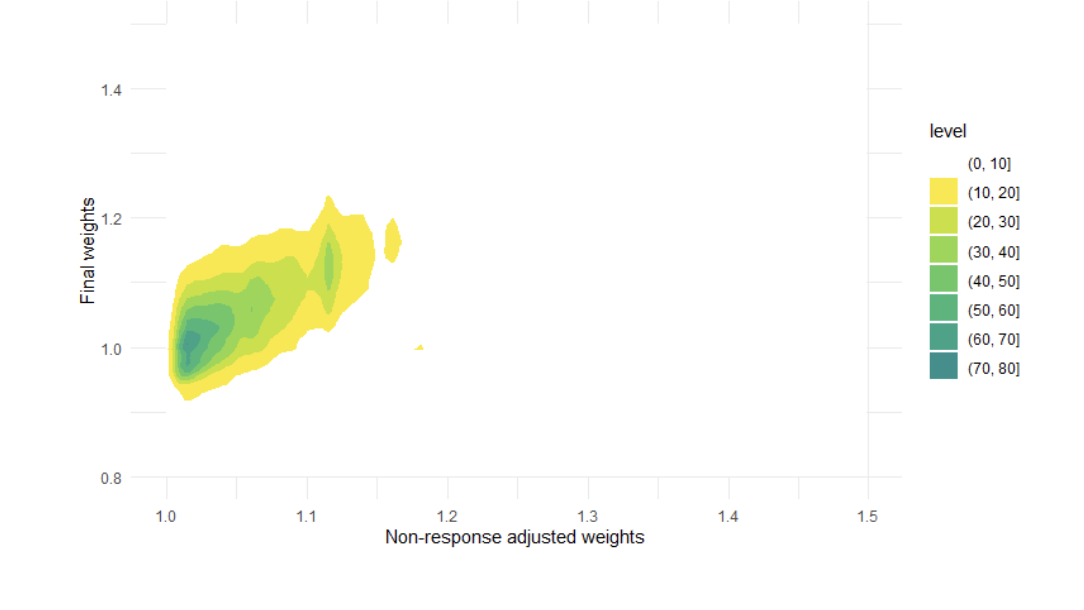}
\caption{Bivariate density of non-response adjusted weights and final calibrated weights.}
\label{fig:bivariate}
\end{figure}

To guarantee that these subsequent demographic adjustments did not introduce erratic design destabilization or artificial inflation into the localized structures, we evaluate the joint behavior of both expansion stages. Figure~\ref{fig:bivariate} displays the bivariate density mapping the initial non-response adjusted weights against the final calibrated weights. The tight concentration of the mass along the main diagonal confirms that the second-stage calibration leaves most weights nearly unchanged while smoothly modifying the targeted underrepresented subgroups.

\subsection{Edge Cases and Quality Control}

Three practical issues require dedicated treatment. \emph{Certainty units} are assigned a weight of one because no auxiliary information justifies inflating them or because of their institutional nature: people experiencing homelessness, enumerated through the 2023 point-in-time count by the Ministry of Social Development, and residents of large collective dwellings (i.e., 50 or more residents, such as correctional facilities), which are few, concentrate many individuals, and are most appropriately taken to represent only themselves. \emph{Weights below one} arise naturally under a chi-square distance (equivalently, a linear-model GREG) when demographic groups overrepresented in the field must be down-weighted to match the reference structure; this does not compromise the asymptotic properties of the estimators. \emph{Locality-level inconsistencies} were addressed, prior to calibration, by a ratio (H\'ajek) adjustment treating localities as estimation domains, benchmarking weighted locality sizes against external sources (i.e., the previous census dwelling counts, spatial indicators, and residential electricity meters from the state utility UTE, whose near-universal grid coverage makes its customer registry a reliable proxy for occupied dwellings), with bounded corrections relative to the 2011 benchmarks \citep{brackstone1987,rao2015}.
\\

Quality control accompanied each stage. The empirical distribution of the final weights was monitored through the Kish design effect, the mean, standard deviation, range, percentiles, and intra-segment variance; outlying weights triggered an audit of the auxiliary variables in the corresponding segments. Internal consistency checks verified that weighted estimates exactly reproduce the calibration benchmarks (i.e., total population and population by sex, single year of age, and department), ensuring that the analytic microdata align with the official demographic structure.

\section{Impact on Census Indicators}

To assess the framework's effect, a set of social and demographic indicators was compared before and after weighting. The adjustment consistently increases the representation of subgroups that exhibited higher omission, particularly those in vulnerable contexts and the age-sex groups with lower response propensity.
\\

Table~\ref{tab:indicators} summarizes illustrative indicators, comparing the combined-census figure with the weighted-census estimate and reporting the estimated share of the omitted population accounted for by each attribute. Across indicators, the weighted estimates are higher than the combined-census figures, and the gap is largest for attributes most associated with vulnerability. Residence in informal settlements rises from 4.5\% to 5.5\%, and the rural population from 4.1\% to 5.0\%; housing-deprivation indicators such as inadequate materials, overcrowding, and the absence of appropriate cooking space all increase. The pattern is coherent with the PES diagnosis: correcting differential undercoverage raises measured deprivation, precisely because the omitted population was disproportionately deprived.

\begin{table}[H] 
\centering
\scriptsize 
\renewcommand{\arraystretch}{0.85} 
\setlength{\tabcolsep}{4pt} 

\caption{Selected indicators: combined census versus weighted census, with the estimated share of each attribute in the omitted population.}
\label{tab:indicators}
\begin{tabular}{@{}p{4.6cm}rrrrrr@{}}
\toprule
Indicator & CC (n) & CC (\%) & WC (n) & WC (\%) & Diff. & Omit. (\%) \\
\midrule
Inadequate dwelling materials       & 534,561 & 17.3 & 622,165 & 18.1 & 87,604 & 24.3 \\
Overcrowding ($>$2 persons/room)    & 416,066 & 13.5 & 489,576 & 14.3 & 73,510 & 20.4 \\
Overcrowding ($>$3 persons/room)    & 113,055 & 3.7  & 134,824 & 3.9  & 21,769 & 6.0  \\
Adequate cooking space (deficit)    & 63,359  & 2.1  & 74,809  & 2.2  & 11,450 & 3.2  \\
Heating                             & 364,408 & 11.8 & 417,051 & 12.2 & 52,643 & 14.6 \\
Food preservation                   & 109,191 & 3.5  & 126,913 & 3.7  & 17,722 & 4.9  \\
Water heater for bathing            & 196,666 & 6.4  & 231,872 & 6.8  & 35,206 & 9.7  \\
Drinking-water access               & 135,124 & 4.4  & 156,892 & 4.6  & 21,768 & 6.0  \\
Sanitation access/quality           & 142,443 & 4.6  & 167,657 & 4.9  & 25,214 & 7.0  \\
Residence in informal settlement    & 158,727 & 4.5  & 193,260 & 5.5  & 34,533 & 9.6  \\
Rural population                    & 142,745 & 4.1  & 174,765 & 5.0  & 32,020 & 9.2  \\
\bottomrule
\end{tabular}

\begin{minipage}{\linewidth}
{\tiny \textit{Note:} CC = combined census; WC = weighted census; Diff. = absolute difference (WC count minus CC count); Omit. (\%) = estimated share of the attribute in the omitted population.}
\end{minipage}
\end{table}

\section{Variance Estimation and Quality Measures}

Under design-based inference, estimate quality is summarized by the standard error (SE). For a non-probability sample the relevant criterion is the mean squared error, which adds the residual bias,

\begin{equation}
\mathrm{MSE}(\hat{\theta}) = \mathrm{SE}^2(\hat{\theta}) + \mathrm{B}(\hat{\theta})^2 .
\end{equation}

Residual bias is, in practice, unknown and very difficult to estimate, so precision is approximated from the estimated variance under the assumption that residual bias is a small fraction of the MSE; the framework's role is to make that assumption tenable by minimizing bias through the response and calibration models.
\\

In standard applications, SEs for DR estimators are obtained by resampling (i.e., bootstrap or jackknife). Two features of the census setting make this impractical. First, the non-response adjustment relies on segment-level operational indicators (i.e., the CAWI linkage rate) that act as predefined parameters rather than re-estimated probabilistic models, so the variance contributed by this stage is stable and resampling adds little. Second, with more than three million records, resampling is computationally prohibitive for production and for downstream use by external analysts. We therefore approximate the variance through an equivalent sampling design that reproduces the main sources of variability: clustering at the household level, a finite-population correction reflecting the very high coverage rate in sampling terms, stratification at the census-segment level with segment-varying response rates, and the additional variability introduced by weight dispersion.
\\

This leads to an equivalent stratified cluster design, evaluated by the ultimate-cluster method \citep{hansen1953,wolter2007}, in which the variability between primary sampling units captures the total design variance:
\begin{equation}
\widehat{\mathrm{SE}}^2(\hat{Y}) = \sum_{c=1}^{C} \frac{m_c\,(1 - \hat{\phi}_c)}{m_c(m_c - 1)} \sum_{i \in s_{B_c}} \bigl(\hat{Y}_{ci} - \hat{\bar{Y}}_c\bigr)^2,
\end{equation}

where $m_c$ is the number of enumerated households in segment $c$, $\hat{\phi}_c$ the estimated sampling fraction, $\hat{Y}_{ci} = \sum_{k \in s_{B_{ci}}} w_{ik} y_{ik}$ the weighted household total, and $\hat{\bar{Y}}_c = m_c^{-1}\sum_{i \in s_{B_c}} \hat{Y}_{ci}$.
\\

From the SE, confidence intervals follow as the point estimate plus or minus a critical value times the SE, and the coefficient of variation as $\mathrm{CV} = (\mathrm{SE}/\text{estimate}) \times 100$. The CV supports comparison across estimates of different magnitudes but inflates artificially for very low-prevalence indicators, for which precision should be judged jointly with the SE and the interval width. INE classifies estimates for dissemination by CV bands (excellent, below 5\%; very good, 5\% to 10\%; good, 10\% to 15\%; acceptable, 15\% to 25\%; use with caution, 25\% to 35\%; do not publish, 35\% or above). Operational guidance for computing these quantities in R is provided as supplementary material.

\section{Assumptions and Limitations}

The framework reduces, rather than eliminates, undercoverage error, and the validity of its estimates rests on assumptions that should be stated plainly.
\\

The response model assumes conditional ignorability: given the auxiliary covariates, enumeration probabilities are homogeneous within a segment and among similar households, and undercoverage is independent of the substantive variables. This may not hold exactly, since unobserved factors can influence both response and outcomes, and residual within-segment heterogeneity may persist. The use of the segment as the adjustment domain reduces spatial heterogeneity, and the subsequent calibration mitigates residual misspecification, but neither guarantees that all bias is removed.
\\

The calibration stage carries an implicit superpopulation model, under which the outcomes are partly explained by age, sex, and department. The stronger that association, the greater the reduction in both bias and variance; where the chosen controls fail to capture differences between the enumerated and omitted populations, the correction is incomplete.
\\

Finally, the framework depends on the quality of the auxiliary information. The CAWI linkage rate, used as a contact proxy, can be confounded by internet penetration, device availability, and modal preferences, and the linkage process can introduce matching error. The calibration benchmarks come from the combined census, whose administrative component varies in coverage with institutional interaction; although the demographic totals are considered reliable, they cannot be guaranteed free of measurement error, and errors in the benchmarks propagate into the weights. Violations of the LRBM assumptions, measurement error in the covariates, or misspecified reference totals can therefore degrade the weights and the resulting estimates.

\section{Discussion and Conclusions}

The 2023 Uruguayan census exemplifies a problem that national statistical offices increasingly face: non-response large enough, and differential enough, to compromise the representativeness of the enumerated data. Recovering the aggregate count through administrative integration is necessary but insufficient, because it leaves the substantive variables uncorrected. Treating the enumerated households as a non-probability sample and weighting them with a doubly robust estimator addresses the substantive problem directly. The estimator's defining property, namely consistency when either the response model or the superpopulation model is correct, is precisely what one wants when the selection mechanism is unknown, and it offers more stability than approaches that rely on a single model, on single-stage imputation, or on univariate calibration alone.
\\

Two design choices were central. Modeling response at the census-segment level, using the web linkage rate as a contact proxy, captured spatial heterogeneity that broad-domain corrections miss and was validated against the PES. Calibrating to combined-census totals by age, sex, and department imported reliable demographic structure while deliberately avoiding the finer geography most affected by administrative allocation error. Implemented at the scale of more than three million records, with explicit treatment of certainty units, sub-unit weights, and locality-level inconsistencies, the framework yields weighted estimates that move social indicators in the direction and magnitude implied by the independent PES diagnosis.
\\

The approach has clear limits, set out in Section~7, and it should be read as a bias-reduction strategy rather than a guarantee of unbiasedness. From the user's standpoint, this implies a firm rule: the weighted microdata are the baseline for any population estimate, and unweighted counts should not be used for indicator production. More broadly, the case suggests that techniques developed for non-probability inference (i.e., doubly robust estimation, response-propensity modeling, and calibration), combined with the integration of multiple auxiliary sources, constitute a viable paradigm for official statistics under the coverage pressures that modern censuses now routinely encounter.

\end{document}